\documentclass[pdflatex,sn-mathphys-num]{sn-jnl}

\usepackage{graphicx}%
\usepackage{multirow}%
\usepackage{amsmath,amssymb,amsfonts}%
\usepackage{amsthm}%
\usepackage{mathrsfs}%
\usepackage[title]{appendix}%
\usepackage{xcolor}%
\usepackage{textcomp}%
\usepackage{manyfoot}%
\usepackage{booktabs}%
\usepackage{algorithm}%
\usepackage{algorithmicx}%
\usepackage{algpseudocode}%
\usepackage{listings}%
\usepackage{fullpage}
\usepackage{xcolor} 
\newcommand*\update[1]{\textcolor{black}{#1}}

\theoremstyle{thmstyleone}%
\theoremstyle{thmstyletwo}%

\theoremstyle{thmstylethree}%

\begin{document}

\title[Article Title]{Laminar boundary layers over small-scale textured surfaces}


\author*[1,2]{\fnm{Samuel D.} \sur{Tomlinson}}\email{samuel.tomlinson@greenwich.ac.uk}

\author[3]{\fnm{Demetrios T.} \sur{Papageorgiou}}\email{d.papageorgiou@imperial.ac.uk}
\equalcont{These authors contributed equally to this work.}

\affil*[1]{\orgdiv{School of Computing and Mathematical Sciences}, \orgname{University of Greenwich}, \orgaddress{\city{London}, \postcode{SE10 9LS}, \country{UK}}}

\affil[2]{\orgdiv{Department of Engineering}, \orgname{University of Cambridge}, \orgaddress{\city{Cambridge}, \postcode{CB2 1PZ}, \country{UK}}}

\affil[3]{\orgdiv{Department of Mathematics}, \orgname{Imperial College London}, \orgaddress{\city{London}, \postcode{SW7 2AZ}, \country{UK}}}


\abstract{
We develop a model for steady, laminar boundary layers over small-scale textured surfaces. 
Although the texture is small relative to the boundary-layer thickness, it modifies the flow via a slip length. 
We use matched asymptotic expansions to simplify the problem, dividing the flow into outer, boundary-layer and inner regions. 
The far-field behaviour of the inner problem yields a slip boundary condition for the boundary layer. 
We derive an asymptotic solution valid when the slip length is small, and for arbitrary slip lengths, we develop a numerical method combining Chebyshev collocation and finite differences. 
\update{We apply this framework to canonical small-scale textured surfaces, including superhydrophobic surfaces and riblets, and utilise existing analytical slip formulae. 
However, the framework is expected to extend to liquid-infused, porous, compliant or deformable surfaces with a variety of regular or random textures. 
We demonstrate how slip modifies the boundary layer’s velocity field, wall shear stress and displacement thickness across a range of surface configurations, and examine the linear stability of the resulting slip-modified boundary layers. 
Our approach enables computationally inexpensive modelling of a wide range of small-scale textured surfaces within laminar boundary-layer flows, providing predictive capability for drag, boundary-layer growth and transition across applications ranging from microfluidics to turbo-machinery and marine transport. 
}
}




\maketitle

\section{Introduction}

In high-Reynolds-number flow, a thin boundary layer exists near the wall where viscous effects are significant, even though inertia dominates the bulk flow. 
This fundamental structure, formalised by Prandtl in 1904, enables simplification of the Navier--Stokes equations via matched asymptotic expansions, separating the viscous near-wall region from the outer inviscid flow~\cite{prandtl1905uber}. 
Within this viscous near-wall region, normal diffusion balances streamwise advection, allowing the flow to be described by a reduced model: the boundary-layer equations. 
For steady, incompressible flow over a flat plate, the boundary-layer thickness scales as \(\hat{\delta}_{99} \sim (\hat{\nu} \hat{x}/\hat{U})^{1/2}\), where $\hat{\nu}$ is the kinematic viscosity, $\hat{x}$ the streamwise coordinate and $\hat{U}$ the free-stream velocity~\cite{schlichting2017boundary}. 
Downstream, the boundary layer grows in thickness as the velocity adjusts from the free-stream value $\hat{U}$ to the no-slip condition at the wall. 
For a uniform free-stream velocity over a smooth, infinitely long flat plate, the Blasius solution emerges, providing an exact similarity reduction of the boundary-layer equations to an ordinary differential equation (ODE)~\cite{blasius1907grenzschichten}. 

The classical boundary-layer framework assumes a smooth, impermeable surface and a no-slip condition at the wall. 
However, many natural or engineered surfaces, such as superhydrophobic surfaces (SHSs), liquid-infused surfaces (LISs), riblets and porous surfaces, exhibit texture that can modify the viscous near-wall flow~\cite{lauga2003effective, hardt2022flow, bechert1997experiments, feuillebois2009effective, ponomarev2003surface, beavers1967boundary}. 
These textures typically have a characteristic normal height $A$ and streamwise half-period $\epsilon$, which may be small relative to the boundary-layer thickness $\delta_{99}$, yet their cumulative effect can be significant, as they alter near-wall shear and flow resistance, ultimately influencing the boundary-layer flow~\cite{bottaro2019flow}. 
When a separation of scales exists (i.e., $\epsilon \ll \delta_{99}$), the small-scale texture can be homogenised, meaning its effect is captured through an effective boundary condition for the boundary-layer problem, rather than being explicitly resolved. 
A common effective boundary condition in this context is a slip boundary condition, characterised by a slip length $\lambda$, which encapsulates the geometry and flow properties of the small-scale texture by modulating the velocity at the wall~\cite{lauga2003effective, hardt2022flow, bechert1997experiments, feuillebois2009effective, ponomarev2003surface, beavers1967boundary}. 
The effect of slip on the boundary layer depends on the relative magnitude of $\lambda$ and $\delta_{99}$~\cite{schlichting2017boundary, tomlinson2021superhydrophobic}. 
When $\lambda \ll \delta_{99}$, slip acts as a small perturbation to the classical boundary layer, and Blasius' solution remains valid to leading order. 
When $\lambda \sim \delta_{99}$, slip modifies the classical boundary-layer structure, causing Blasius' solution to break down. 
When $\lambda \gg \delta_{99}$, homogenisation of the surface fails, and the texture must be resolved directly. 

Flow over a small-scale textured surface divides into three regions, reflecting the balance between inertial and viscous forces and forming the basis for constructing matched asymptotic solutions~\cite{prandtl1905uber, tomlinson2021superhydrophobic}. 
The outer region, governed by the Euler equations, is inviscid and unaffected by the wall. 
Viscous effects are negligible here, and the velocity field is uniform, reflecting the dominant inertial dynamics in the bulk. 
The middle region corresponds to the boundary layer, where inertial and viscous forces are both significant and balanced. 
Here, the velocity adjusts to match the outer inviscid flow and inner viscous region. 
The boundary layer governs the wall shear stress responsible for drag. 
The inner region lies closest to the wall, with a thickness of $O(\epsilon)$, corresponding to the small-scale texture~\cite{luchini1991resistance, tomlinson2021superhydrophobic}. 
In the inner region, inertial effects are negligible compared to viscous forces, and the texture must be fully resolved. 
The flow is governed by the Stokes equations, and its far-field behaviour determines the slip length $\lambda$ in the boundary layer. 
Using matched asymptotic expansions, these three regions can be connected to construct a solution valid across all scales, allowing the small-scale texture to be replaced by a slip boundary condition in the boundary-layer equations. 

In particular, a variety of small-scale textured surfaces have been investigated using multi-scale modelling and experiments~\cite{bottaro2019flow}. 
On superhydrophobic surfaces (SHSs), slip arises from gas pockets trapped between ridges or pillars~\cite{joseph2006slippage}. 
The slip length for SHSs depends on the small-scale texture geometry, gas fraction and alignment relative to the flow direction. 
Analytical and semi-analytical models describing SHS slip behaviour include those by Philip~\cite{philip1972flows}, Lauga and Stone~\cite{lauga2003effective} and Ybert et al.~\cite{ybert2007achieving}. 
Liquid-infused surfaces (LISs), in which the small-scale texture is filled with a second immiscible fluid, exhibit slip lengths that depend not only on geometry but also on viscosity contrast, interfacial tension and capillary effects (Wong et al.~\cite{wong2011bioinspired}). 
Studies by Hardt and McHale~\cite{hardt2022flow}, Sundin and Bagheri~\cite{sundin2022slip} and Van Buren and Smits~\cite{van2017substantial} have extended SHS-style analyses to LISs. 
Riblets have long been employed for turbulent drag reduction and can also produce slip effects in laminar flows. 
Their influence can be anisotropic, which is modelled using slip tensors that account for the alignment of the riblets relative to the flow~\cite{luchini1991resistance, bechert1997experiments, bechert2000experiments, walsh1983riblets}. 
Other periodic textures have been analysed using periodic homogenisation, yielding expressions for slip in Feuillebois et al.~\cite{feuillebois2009effective}, Zhou et al.~\cite{zhou2013effective}, and Sharma et al.~\cite{sharma2020influence}. 
For random textures, stochastic homogenisation and empirical fits to experiments have been used to develop models for slip~\cite{sbragaglia2007effective, ponomarev2003surface, priezjev2011molecular, cottin2012scaling}. 
Porous surfaces can be modelled using the Beavers--Joseph condition, which gives a slip length proportional to the square root of the permeability~\cite{beavers1967boundary, nield2009beavers}. 

In this work, we develop a model for steady, two-dimensional (2D) laminar boundary layers over small-scale textured surfaces. 
Assuming the texture is small compared to the boundary layer thickness, we homogenise the surface's effect into a slip length $\lambda$~\cite{bottaro2019flow}, resulting in a slip boundary condition of the form  $u = \lambda u_y$, which modifies Blasius' solution~\cite{tomlinson2021superhydrophobic}. 
\update{We derive an asymptotic solution valid in the small-slip limit $\lambda \ll 1$, which yields explicit scaling relations for the changes in displacement thickness and wall shear stress and provides a benchmark for verifying the numerical boundary-layer solver (and vice versa). 
We construct a numerical solution for $\lambda = O(1)$ using Chebyshev collocation in the normal direction and finite differences in the streamwise direction~\cite{trefethen2000spectral}.} 
This asymptotic--numerical approach efficiently and accurately resolves flows for arbitrary slip lengths.
\update{Crucially, this allows surface-induced slip effects to be incorporated into boundary-layer models for a range of laminar flows, enabling prediction of boundary-layer thickness and drag relevant to analysis and design, without performing texture-resolved simulations, thereby reducing computational cost. 
In addition to characterising the steady boundary-layer structure, we examine the linear stability of the resulting slip-modified boundary layers, demonstrating how surface texture influences stability characteristics within this framework. 
Applications of this model include drag-reduction and flow-control technologies across multiple length and Reynolds-number scales, ranging from microfluidic devices to turbo-machinery blades and marine transport operating in laminar flow regimes. 
We demonstrate the generality of this approach by applying it to representative small-scale periodic textures with known analytical slip lengths, SHSs and riblets, which serve as canonical examples for the broader modelling framework and for potential applications to other textured surfaces, including LISs, porous, compliant and deformable surfaces. 
}

The remainder of the paper is organised as follows. 
In Section~\ref{sec:formulation}, we formulate the problem of flow over small-scale textured surfaces. 
Section~\ref{sec:model} simplifies the problem to a 2D laminar boundary layer with a slip boundary condition. 
In Section~\ref{sec:methods}, we derive the asymptotic solution for small slip lengths and describe the numerical solution for arbitrary slip lengths. 
\update{Section~\ref{sec:stability} presents a linear stability analysis of the slip-modified boundary layer.} 
Section~\ref{sec:results} analyses boundary-layer solutions for several small-scale textured surfaces. 
Finally, Section~\ref{sec:discussion} summarises the results, highlights applications of this framework, and outlines potential extensions. 

\section{Formulation}
\label{sec:formulation}

\subsection{Governing equations} 

\begin{figure*}[t!]
    \centering
    (a) \hfill (b) \hfill \vspace{.12cm} \\
    \includegraphics[width=\linewidth]{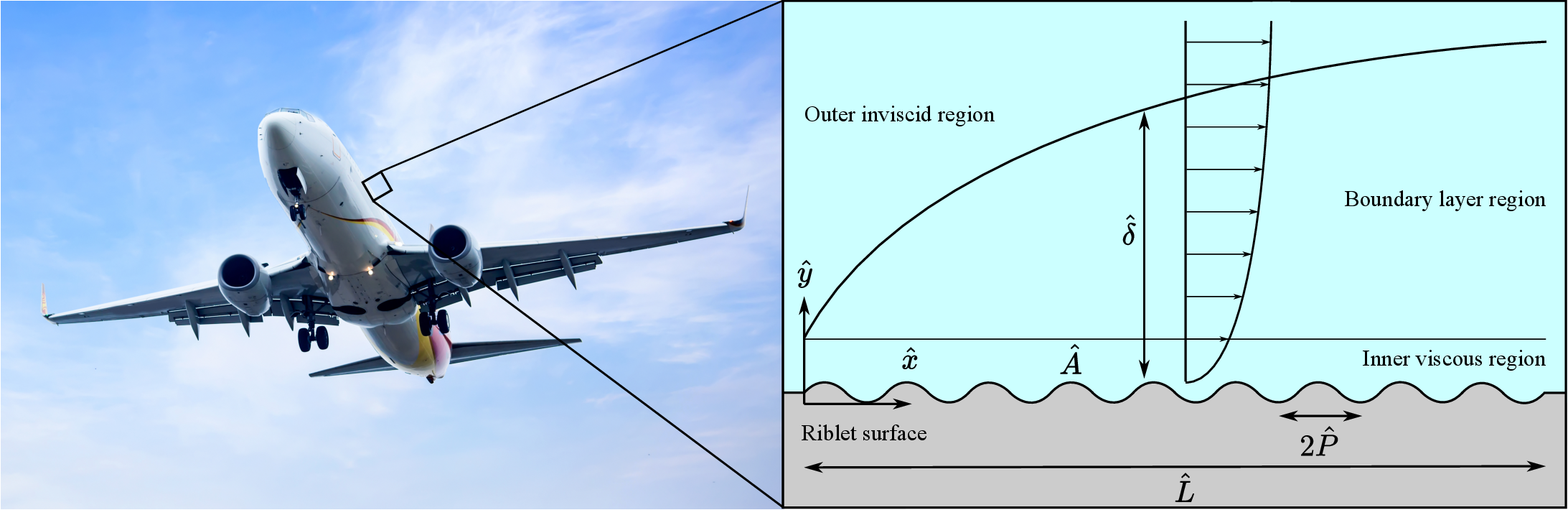}\\ 
    (c) \hfill (d) \hfill \vspace{.12cm} \\
    \includegraphics[width=\linewidth]{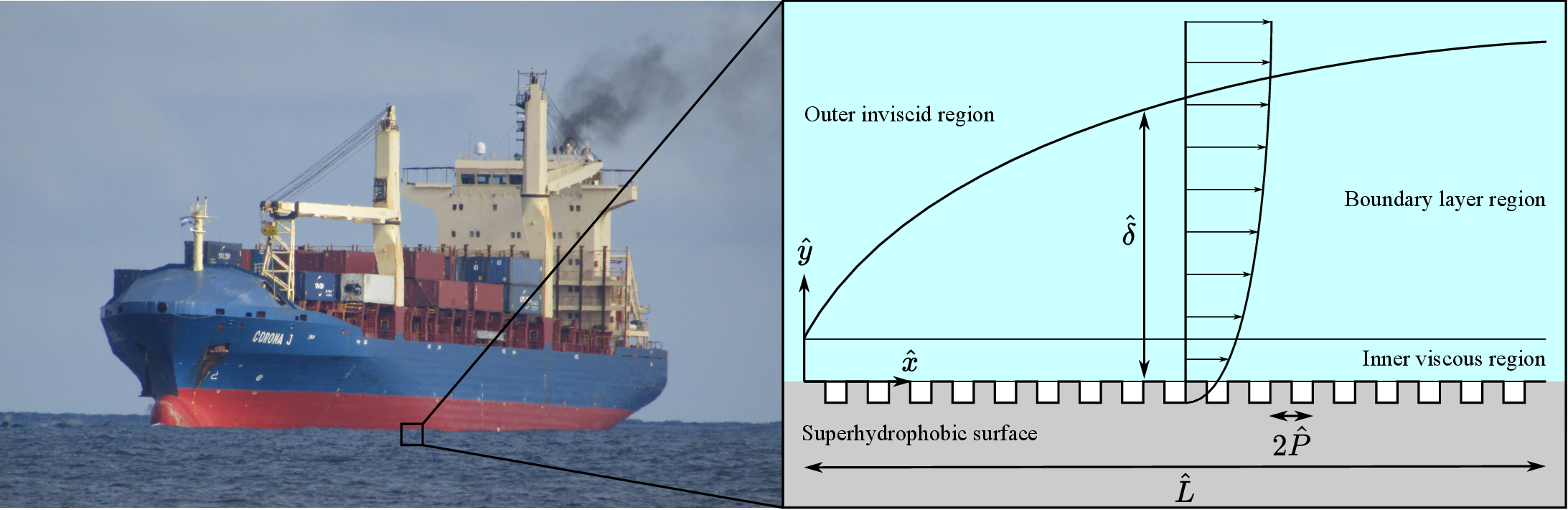}\\ 
    \caption{(a) A ship with a potential superhydrophobic coating. 
    (b) Schematic of the boundary layer forming over a transversely ridged SHS with texture half-period $\hat{P}$ and \update{reference length} $\hat{L}$. 
    (c) An aeroplane that could be equipped with riblets. 
    (d) Schematic of the boundary layer forming over a transversely ribleted surface with texture amplitude $\hat{A}$, texture half-period $\hat{P}$ and reference length $\hat{L}$. 
    Schematics in (b) and (d) illustrate the outer inviscid region, boundary layer and inner viscous region. 
    }
    \label{fig:1}
\end{figure*}

We consider 2D laminar flow over a textured surface, such as SHSs or riblets, characterised by a streamwise texture period $2\hat{P}$ and a \update{reference length $\hat{L}$, such that the surface may be finite or effectively infinite} (Figure~\ref{fig:1}). 
A scale separation is assumed between the streamwise texture half-period and \update{reference length}, i.e., $\hat{P}\ll\hat{L}$. 
Dimensional variables are denoted with a hat to distinguish them from the non-dimensional variables introduced in Section~\ref{sec:Non-dimensionalisation and scalings}. 
The spatial coordinates are $\hat{x}$ (streamwise) and $\hat{y}$ (normal) \update{and $\hat{t}$ is time.} 
The velocity components are $\hat{u} = \hat{u}(\hat{x},\,\hat{y}\update{,\,\hat{t}})$ (streamwise) and $\hat{v} = \hat{v}(\hat{x},\,\hat{y}\update{,\,\hat{t}})$ (normal), and the pressure field is $\hat{p} = \hat{p}(\hat{x},\,\hat{y}\update{,\,\hat{t}})$. 
The fluid is assumed incompressible and Newtonian with viscosity $\hat{\mu}$. 
For SHSs, the wall at $\hat{y} = 0$ consists of alternating solid ridges of length $2(1-\phi)\hat{P}$ and liquid--gas interfaces of length $2\phi \hat{P}$, where $0 < \phi < 1$ is the gas fraction. 
The liquid--gas interface is assumed flat, a common simplification in theoretical studies of SHSs that enables analytical treatment of the inner problem. 
In practice, however, interfacial curvature can arise due to pressure gradients or surface tension, which may alter the slip length~\cite{lee2016superhydrophobic}. 

The incompressible Navier--Stokes equations governing the velocity and pressure fields are 
\begin{align} 
\label{eq:dim_1}
    \hat{u}_{\hat{x}} + \hat{v}_{\hat{y}} &= 0, \\
    \update{\hat{u}_{\hat{t}}} +  \hat{u}\hat{u}_{\hat{x}} + \hat{v}\hat{u}_{\hat{y}} + \hat{p}_{\hat{x}}/\hat{\rho} &= \hat{\nu} (\hat{u}_{\hat{x}\hat{x}} + \hat{u}_{\hat{y}\hat{y}}), \\
    \update{\hat{v}_{\hat{t}}} +  \hat{u}\hat{v}_{\hat{x}} + \hat{v}\hat{v}_{\hat{y}} +  \hat{p}_{\hat{y}}/\hat{\rho} &= \hat{\nu} (\hat{v}_{\hat{x}\hat{x}} + \hat{v}_{\hat{y}\hat{y}}),
\end{align}
where $\hat{\rho}$ is the fluid density, $\hat{\nu} = \hat{\mu}/\hat{\rho}$ is the kinematic viscosity and $\hat{\mu}$ is the dynamic viscosity. 
On the liquid--gas portions of the SHS, located at $\hat{x} \in [-\phi \hat{P}, \,\phi \hat{P}] + 2n\hat{P}$ at $\hat{y} = 0$, the tangential stress-free and no-penetration boundary conditions apply: 
\begin{equation} 
\label{eq:dim_2}
    \hat{u}_{\hat{y}} = 0, \quad \hat{v} = 0,
\end{equation}
for all integers $n \geq 0$. 
On the solid ridge portions of the wall, corresponding to $\hat{x} \in [\phi \hat{P}, \,(2 - \phi) \hat{P}] + 2n \hat{P}$ at $\hat{y} = 0$, the no-slip and no-penetration conditions hold: 
\begin{equation} 
\label{eq:dim_3}
    \hat{u}= 0, \quad \hat{v} = 0.
\end{equation}
For riblets, the wall follows a periodic profile $\hat{y} = \hat{A} \sin$($\pi \hat{x}/\hat{P}$), where $\hat{A}$ is the riblet amplitude. 
On this geometry, the standard no-slip and no-penetration boundary conditions given in \eqref{eq:dim_3} are applied. 
As $\hat{y} \to \infty$, the influence of the wall vanishes, and the velocity field approaches the imposed free-stream flow. 
In this work, the flow is driven by a constant velocity $\hat{U}$ imposed far from the wall, so that 
\begin{equation} 
\label{eq:dim_4}
    \hat{u} \to \hat{U}, \quad \hat{v} \to 0,
\end{equation}
as $\hat{y} \rightarrow \infty$. 

\subsection{Non-dimensionalisation and scalings} \label{sec:Non-dimensionalisation and scalings}

To simplify the problem and identify the relevant dimensionless groups, we non-dimensionalise \eqref{eq:dim_1}--\eqref{eq:dim_4} using the characteristic velocity $\hat{U}$ and \update{reference length} $\hat{L}$. 
We introduce the following non-dimensional variables: 
\begin{equation} 
\label{eq:non_dim_1} 
    \update{t = \hat{t}/(\hat{L}/\hat{U}),} \quad x = \hat{x}/\hat{L}, \quad y = \hat{y}/\hat{L}, \quad u = \hat{u}/\hat{U}, \quad v = \hat{v}/\hat{U}, \quad p = \hat{p}/(\hat{\rho}\hat{U}^2),
\end{equation}
and define the dimensionless texture half-period 
\begin{equation} 
\label{eq:non_dim_2}
    \epsilon = \hat{P}/\hat{L},
\end{equation}
which we assume satisfies $\epsilon \ll 1$. 
The small parameter $\epsilon$ quantifies the ratio of texture to flow scale and plays a central role in the asymptotic analysis that follows. 
Substituting the non-dimensional variables \eqref{eq:non_dim_1}--\eqref{eq:non_dim_2} into the dimensional Navier--Stokes equations \eqref{eq:dim_1}, we obtain the non-dimensionalised system: 
\begin{align} 
\label{eq:nondim_1}
    u_{x} + v_{y} &= 0, \\
    \text{Re}(\update{u_t} + uu_{x} + vu_{y} + p_{x}) &= u_{xx} + u_{yy}, \\
    \label{eq:nondim_1c}
    \text{Re}(\update{v_t} + uv_{x} + vv_{y} + p_{y}) &= v_{xx} + v_{yy},
\end{align}
where $\text{Re} = \hat{U} \hat{L} / \hat{\nu}$ is the Reynolds number, characterising the ratio of inertial to viscous effects. 
On the liquid--gas interface of the SHS at $x \in [-\phi \epsilon, \, \phi \epsilon] + 2n \epsilon$ and $y = 0$, we impose 
\begin{equation} 
\label{eq:nondim_2}
    u_{y} = 0, \quad v = 0,
\end{equation}
while on the solid portions at $x \in [\phi \epsilon, \, (2 - \phi)\epsilon] + 2n \epsilon$ and $y = 0$, we impose 
\begin{equation} 
\label{eq:nondim_3}
    u = 0, \quad v = 0.
\end{equation}
For riblets, with surface $y = A \sin(\pi x / \epsilon)$, where $A = \hat{A} / \hat{L}$ is the non-dimensional amplitude, the no-slip and no-penetration conditions \eqref{eq:nondim_3} are applied along the riblet surface. 
\update{For the purposes of the stability analysis performed in Section~\ref{sec:stability}, the boundary conditions for the SHS or riblets can be approximated as an effective condition with slip
\begin{equation} \label{eq:full_slippy}
    u = \epsilon \beta u_{y}, \quad v=0,
\end{equation}
where $\beta$ is the slip length~\cite{lauga2003effective, hardt2022flow, bechert1997experiments, feuillebois2009effective, ponomarev2003surface, beavers1967boundary}. 
}
Far from the wall, as $y \to \infty$, the flow approaches the imposed uniform stream: 
\begin{equation} 
\label{eq:nondim_4}
    u \to 1, \quad v \to 0.
\end{equation} 
The non-dimensionalised system \eqref{eq:nondim_1}--\eqref{eq:nondim_4}, with $\epsilon \ll 1$ and $\text{Re} \gg 1$ as the relevant small and large parameters, provides the starting point for the asymptotic analysis developed in the next section. 

\section{Model} \label{sec:model}

To simplify the governing equations \eqref{eq:nondim_1}--\eqref{eq:nondim_4}, we exploit the scale separation between the textured surface and the boundary-layer flow. 
For SHSs, we consider the asymptotic regime $\epsilon \ll \text{Re}^{-1/2} \ll 1$. 
This inequality is justified for specific applications in Section~\ref{subsec:SHSs_riblets}. 
The flow is analysed by decomposing the domain into multiple regions, each governed by a distinct balance of inertial and viscous forces. 
For riblet geometries, we further assume $A \ll \epsilon \ll \text{Re}^{-1/2} \ll 1$, where $A$ is the non-dimensional amplitude of the riblets, ensuring that the riblet height acts as a small perturbation within the textured scale. 
Under these scalings, the flow near the surface can be treated as locally periodic, modulated by the slowly varying boundary-layer shear defined in Section~\ref{sec:middle}. 
The leading-order boundary value problem, derived from \eqref{eq:nondim_1}--\eqref{eq:nondim_4}, is solved using a combination of asymptotic expansions and numerical computations. 
The objective is to determine how small-scale period textures influence the development of the boundary-layer. 

To this end, the flow is decomposed into three distinct regions, each associated with a dominant physical mechanism and a characteristic spatial scale. 
The outer region (Section \ref{sec:outer}), or inviscid layer, is defined by $x=O(1)$ and $y = O(1)$, where viscous effects are negligible. 
The middle region (Section \ref{sec:middle}) corresponding to the boundary layer, is characterised by $x=O(1)$ and $y = O(\text{Re}^{-1/2})$, where viscous--inertial interactions are significant. 
The inner region (Section \ref{sec:inner}), located near the wall and around the small-scale texture, with $x = O(\epsilon)$ and $y = O(\epsilon)$, where viscous effects dominate. 
The governing equations are analysed in each region under the appropriate scaling, solutions are constructed, and then matched asymptotically in the overlapping limits. 
This composite asymptotic solution is consistent across regions and captures the multi-scale structure of the boundary layer. 

\subsection{Outer region (inviscid region)} \label{sec:outer}

In the outer region, the flow is governed by inviscid dynamics, corresponding to the bulk flow where the influence of the wall is negligible and inertia dominates. 
In this region, the spatial coordinates scale as $x = O(1)$ and $y = O(1)$, with velocity components $u = O(1)$ and $v = O(1)$. 
\update{Assuming the flow is steady,} the flow and pressure fields are expanded in powers of $\text{Re}^{-1}$: 
\begin{equation} 
\label{eq:exp_inv}
    x = x^*, \quad y = y^*, \quad u = u^*(x^*,\,y^*) + \dots, \quad v = v^*(x^*,\,y^*) + \dots, \quad p = p^*(x^*,\,y^*) + \dots.
\end{equation}
Substituting the expansion \eqref{eq:exp_inv} into the non-dimensional Navier–Stokes equations \eqref{eq:nondim_1}, assuming the flow is steady and taking the limit $\text{Re} \to \infty$, the leading-order outer equations reduce to the incompressible Euler equations: 
\begin{align} 
\label{eq:outer_1}
    u^*_{x^*} + v^*_{y^*} &= 0, \\
    u^* u^*_{x^*} + v^* u^*_{y^*} + p^*_{x^*} &= 0, \\
    u^* v^*_{x^*} + v^* v^*_{y^*} + p^*_{y^*} &= 0.
\end{align}
At the lower boundary $y^* = 0$, we impose the no-penetration condition from \eqref{eq:nondim_2} and \eqref{eq:nondim_3}, neglecting tangential conditions since viscosity is absent, such that 
\begin{equation} 
\label{eq:outer_2}
    v^* = 0.
\end{equation}
At the upper boundary, the free-stream condition from \eqref{eq:nondim_4} is applied: 
\begin{equation} 
\label{eq:outer_3}
    u^* \to 1,
\end{equation}
as $y^* \to \infty$. 
The solution to \eqref{eq:outer_1}--\eqref{eq:outer_3}  satisfying the free-stream and wall boundary conditions is 
\begin{equation} 
\label{eq:outer_4}
    u^* = 1, \quad v^* = 0.
\end{equation}
This solution represents the undisturbed bulk flow and serves as the far-field condition for the boundary layer flow in the middle region, which in turn defines the slowly varying shear field that drives the inner, locally periodic problems. 

\subsection{Middle region (boundary layer)} \label{sec:middle}

In the middle region, corresponding to the boundary layer, inertia and viscosity are balanced in the normal direction, while streamwise viscous terms are negligible for $\text{Re}\gg 1$. 
We stretch the normal coordinate as 
\begin{equation} 
\label{eq:bl_scale_1}
    y' = \text{Re}^{1/2} y,
\end{equation}
reflecting the classical boundary-layer thickness scaling $\delta_{99} =O( \text{Re}^{-1/2})$ for high-Reynolds number flows~\cite{schlichting2017boundary}. 
\update{Assuming the flow is steady,} we scale and expand the space and velocity fields: 
\begin{equation} 
\label{eq:bl_scale_2}
    x = x', \quad
    u = u'(x',\,y') + \dots, \quad
    v = v'(x',\,y')/\text{Re}^{1/2} + \dots,
\end{equation}
with $u = O(1)$ and $v = O(\text{Re}^{-1/2})$. 
Substituting \eqref{eq:bl_scale_1}--\eqref{eq:bl_scale_2} into the non-dimensional Navier--Stokes equations \eqref{eq:nondim_1} and retaining leading-order terms as $\text{Re} \to \infty$, we obtain the Prandtl boundary layer equations: 
\begin{align} 
\label{eq:middle_1}
    u'_{x'} + v'_{y'} &= 0, \\
    \label{eq:middle_1b}
    u' u'_{x'} + v' u'_{y'} &= u'_{y'y'},
\end{align}
\update{where $p'_{x'} = 0$, as the outer flow is uniform and therefore has no imposed pressure gradient. 
Consequently, the leading-order inner Stokes problem (Section~\ref{sec:inner}) will be driven solely by viscous shear imposed through matching with the boundary-layer velocity field. 
}
These equations describe the dominant physics within the thin viscous layer adjacent to the wall, where the velocity adjusts to satisfy the boundary condition at $y = 0$. 
At the outer edge of the boundary layer, matching to the inviscid region \eqref{eq:outer_4} requires 
\begin{equation} \label{eq:middle_2}
    u' \to 1 \quad \text{as} \quad y' \to \infty.
\end{equation} 

Near the wall, the velocity profile must match the inner problem, which resolves the small-scale geometry of the textured surface. 
The scale separation $\epsilon \ll \text{Re}^{-1/2}$ ensures that the boundary layer does not directly resolve the small-scale surface variations, but instead interacts with the texture through the averaged velocity and shear rate provided by the inner solution. 
As $y' \to 0$, the streamwise velocity can be expanded in a Taylor series: 
\begin{equation} 
\label{eq:taylor_1}
    u'(x', \,y') = u'(x', \,0) + u'_{y'}(x', \,0) y' + \dots,
\end{equation}
where the coefficients vary only on the long streamwise coordinate of the boundary layer. 

\subsection{Inner region (viscous roughness region)} \label{sec:inner}

The inner region corresponds to the small-scale texture, capturing both surface variations and the associated viscous flow. 
We define the stretched coordinates 
\begin{equation} 
\label{eq:inner_1}
    X = x/\epsilon, \quad Y = y/\epsilon,
\end{equation}
so that $X$ and $Y$ are $O(1)$ near the wall. 
\update{Assuming the flow is steady,} the velocity and pressure fields are rescaled and expanded as
\begin{equation} 
\label{eq:inner_exp}
    u = \epsilon U(X,\,Y,\,x',\,y') + \dots, \quad v = \epsilon V(X,\,Y,\,x',\,y') + \dots, \quad p = P(X,\,Y,\,x',\,y')/\text{Re} + \dots.
\end{equation}
where $U$, $V$ and $P$ depend on the short length scales ($X$ and $Y$) and on the long length scales parametrically ($x'$ and $y'$). 
In this regime, viscous forces dominate at leading order, while inertial effects are negligible. 
Substituting the coordinates and expansions \eqref{eq:inner_1}--\eqref{eq:inner_exp} into the non-dimensional Navier-Stokes equations \eqref{eq:nondim_1} yields the incompressible Stokes equations at leading order: 
\begin{align} 
\label{eq:inner_2}
    U_X + V_Y &= 0, \\
    U_{XX} + U_{YY} &= P_X, \\
    V_{XX} + V_{YY} &= P_Y.
\end{align}
Derivatives with respect to the slow variables $x'$ and $y'$ enter \eqref{eq:inner_2} at higher order. 
For SHSs, the boundary conditions at $Y=0$ from \eqref{eq:nondim_2}--\eqref{eq:nondim_3} are:
\begin{equation} 
\label{eq:inner_3}
    U_Y = 0, \quad V = 0, \quad \text{for } X \in [-\phi, \phi] + 2n, \quad U = 0, \quad V = 0, \quad \text{for } X \in [\phi, 2-\phi] + 2n,
\end{equation}
for all integers $n$. 
We assume that the number of cells $N \gg 1$, but the total width $N\epsilon \ll 1$, so that the inner region is contained within the boundary-layer scale.
For riblets with wall profile $Y = A \sin(\pi X) / \epsilon$, we impose the no-slip and no-penetration conditions \eqref{eq:nondim_3} along the wall: 
\begin{equation} 
\label{eq:inner_4}
    U = 0, \quad V = 0.
\end{equation}
For $Y \gg 1$, the inner solution must match the Taylor expansion of the middle-region velocity. 
Based on \eqref{eq:taylor_1}, the far-field inner solution satisfies 
\begin{equation} 
\label{eq:outer_inner_1}
    U \sim \tau (\beta + Y),
\end{equation}
where $\tau = \tau(x', \, y')$ is the local wall shear rate and $\beta = \beta(x', \, y')$ is the slip length. 
Differentiating \eqref{eq:outer_inner_1} gives 
\begin{equation} 
\label{eq:inner_6}
    U_Y \to \tau,
\end{equation}
for $Y \gg 1$. 
Because the surface texture is much smaller than the boundary-layer and outer scales ($\epsilon \ll \text{Re}^{-1/2}\ll 1$), the inner problem is treated as locally periodic in $X$ for each fixed outer coordinate $x'$. 
For certain canonical textures, such as superhydrophobic stripes or sinusoidal riblets, the inner problem \eqref{eq:inner_2}--\eqref{eq:inner_6} can be solved exactly or asymptotically, yielding explicit expressions for $\beta$. 
For SHSs, Philip~\cite{philip1972flows} showed 
\begin{equation} 
\label{eq:philip}
    \beta = \frac{1}{2\pi} \ln \left( \sec \left( \frac{\pi\phi}{2} \right) \right),
\end{equation}
and for riblets, Luchini et al.~\cite{luchini1991resistance} derived 
\begin{equation} 
\label{eq:luchini}
    \beta = -\frac{\pi A^2}{4\epsilon^2},
\end{equation}
where the slip-length expressions from~\cite{philip1972flows} and~\cite{luchini1991resistance} have been converted to the present coordinate system and non-dimensionalisation. 

\subsection{Matching between middle and inner regions}

In the overlapping region, the inner solution ($Y \gg 1$) and the middle solution ($y \to 0$) must asymptotically match. 
The inner expansion \eqref{eq:outer_inner_1} gives 
\begin{equation} 
\label{eq:inner_5}
    u(x,\,y) = \epsilon \tau (\beta + y/\epsilon) + \dots,
\end{equation}
while the middle expansion \eqref{eq:taylor_1} gives 
\begin{equation} 
\label{eq:outer_new}
    u(x,\,y) = u(x,\,0) + u_y(x,\,0) y + \dots.
\end{equation}
Matching these expansions \eqref{eq:inner_5}--\eqref{eq:outer_new} term by term in the overlap region yields 
\begin{equation} 
\label{eq:slip_1}
    u(x,\,0) = \epsilon \beta u_y(x,\,0),
\end{equation}
so that the velocity at the wall is proportional to its normal derivative in a manner prescribed by the inner problem. 
This is the slip condition for laminar boundary layers, where the slip length $\beta$ encodes the small-scale surface effects~\cite{rothstein2010slip}. 
In boundary-layer variables \eqref{eq:bl_scale_1}-\eqref{eq:bl_scale_2}, the slip condition becomes 
\begin{equation} 
\label{eq:matching_1}
    u'(x',\,0) = \lambda u'_{y'}(x',\,0) = \epsilon \beta \text{Re}^{1/2} u'_{y'}(x',\,0),
\end{equation}
which provides the final boundary condition needed to close the boundary-layer equations \eqref{eq:middle_1}–\eqref{eq:middle_2}. 
We now proceed to solve the boundary-layer problem with a slip condition (\ref{eq:middle_1}, \ref{eq:middle_1b}, \ref{eq:middle_2}, \ref{eq:matching_1}) using asymptotic ($\lambda \ll 1$) and numerical methods ($\lambda = O(1)$). 

\section{Methods} 
\label{sec:methods}

\subsection{Numerical solution} 
\label{sec:numerics}

We solve the steady, incompressible, 2D boundary-layer equations subject to a slip boundary condition, as specified in (\ref{eq:middle_1}, \ref{eq:middle_1b}, \ref{eq:middle_2}, \ref{eq:matching_1}). 
\update{Although (\ref{eq:middle_1}, \ref{eq:middle_1b}, \ref{eq:middle_2}, \ref{eq:matching_1}) were derived under the scale separation \(\epsilon \ll \text{Re}^{-1/2}\), which ensures that the inner texture region remains contained within the boundary layer, we also solve these equations numerically to (i) verify the asymptotic predictions in the regime where the ordering implies $\lambda \ll 1$, and (ii) explore the qualitative response of the boundary layer to stronger slip within the same homogenised model. 
Furthermore, while the geometric scale separation \(\epsilon \ll \text{Re}^{-1/2}\) is maintained, the slip length \(\beta\) may become large for certain textures (e.g., SHSs as \(\phi \to 1\)), so that the combined parameter $\lambda = \epsilon \beta \text{Re}^{1/2}$ can be $O(1)$. 
}
To simplify the discretisation and enhance stability, we introduce an auxiliary variable \(q'=\partial u' / \partial x'\), reducing the system to first-order in the streamwise direction for numerical integration. 
Chebyshev spectral collocation is used in the normal direction $y'$~\cite{trefethen2000spectral}, while integration in $x'$ is performed using an implicit marching scheme. 

The normal domain \(y' \in [0, \, y'_\infty]\) is linearly mapped to the standard interval \(\xi \in [-1, \, 1]\), facilitating the use of Chebyshev collocation points and differentiation matrices~\cite{trefethen2000spectral}. 
The height $y'_\infty$ is chosen to capture the full boundary-layer development, ensuring that numerically $u' \rightarrow 1$ as $y' \rightarrow \infty$ consistent with \eqref{eq:middle_2}. 
At each streamwise location $x'_i$, the discrete solution vector contains the values of $\mathbf{u}'$, $\mathbf{v}'$ and $\mathbf{q}'$ at $N_\xi + 1$ collocation points, arranged as $\mathbf{x}' = [\mathbf{u}' \,\,\, \mathbf{v}' \,\,\, \mathbf{q}']^T \in \mathbb{R}^{3(N_\xi+1)}$. 
Spectral differentiation matrices \(D_\xi\) and \(D^2_\xi\) are constructed using standard Chebyshev-collocation methods and scaled to the physical domain~\cite{trefethen2000spectral}. 
At each streamwise step \(x'_i\), the boundary-layer equations (\ref{eq:middle_1}, \ref{eq:middle_1b}, \ref{eq:middle_2}, \ref{eq:matching_1}) are discretised to form a linear system: 
\begin{equation} 
\label{eq:numerical_sol}
\mathbf{A} \, \mathbf{x}' = \mathbf{b}, \quad \text{where} \quad \mathbf{A} =
\begin{bmatrix}
\mathbf{L}_1 & \mathbf{L}_2 & \mathbf{L}_3 \\
\mathbf{M}_1 & \mathbf{M}_2 & \mathbf{M}_3 \\
\mathbf{N}_1 & \mathbf{N}_2 & \mathbf{N}_3
\end{bmatrix} \in \mathbb{R}^{3(N_\xi+1)\times 3(N_\xi+1)},
\end{equation}
where each submatrix \(\mathbf{L}_i,\, \mathbf{M}_i, \,\mathbf{N}_i\) encodes diffusive, advective and continuity contributions for the respective velocity components. 
Boundary conditions are applied by directly replacing the corresponding rows of \(\mathbf{A}\) and entries of \(\mathbf{b}\). 
At the wall (\(y' = 0\)), we impose the slip condition \(u' = \lambda u'_y\) and no penetration \(v' = 0\); at the outer boundary (\(y' = y'_\infty\)), the velocity is set to match the free stream \(u' = 1\). 
Nonlinear terms, such as \(u' \partial u' / \partial y'\), are treated using a fixed-point iteration, where the previous iteration’s solution is used to linearise the system. 
The iteration proceeds until the relative residual, defined as $\tau = \left\| \mathbf{A}(\mathbf{x}'^{(n)} - \mathbf{x}'^{(n+1)}) \right\|_1/r$, falls below a prescribed tolerance of \(10^{-4}\), with a relaxation factor of $r = 0.5$ applied to improve convergence stability. 
Based on convergence tests, we set \(y'_\infty = 30\) and use \(N_\xi = 30\) collocation points in the normal direction and \(N_x = 1000\) streamwise steps over $x'\in[x_0', \, 30]$. 
The resulting scheme combines spectral accuracy in $y'$ with implicit integration in $x'$, providing an efficient, stable and accurate solution for boundary layers with arbitrary slip lengths. 
\update{The downstream marching is not initiated at \(x'=0\), where the boundary-layer equations are singular because the boundary-layer thickness vanishes at the leading edge. 
Instead, the computation is started at a small finite location \(x'_0 \ll 1\). 
In the present computations, $x'_0 = 0.03$. 
The velocity field is initialised at $x'_0$ using the Blasius similarity solution, and the results were verified to be insensitive to the choice of $x'_0$. 
}

\subsection{Key quantities of interest} 

Key quantities of interest are the displacement thickness \(\delta(x')\) and the wall shear stress \(\tau(x')\)~\cite{schlichting2017boundary}. 
The displacement thickness is evaluated from the velocity field using Clenshaw--Curtis quadrature based on Chebyshev collocation weights~\cite{trefethen2000spectral}. 
These quantities characterise how texture-induced slip alters the boundary-layer structure and wall shear, providing key indicators of surface performance~\cite{schlichting2017boundary}. 
The displacement thickness represents the distance by which the effective wall position must be shifted to conserve the free‐stream mass flux, and is defined as 
\begin{equation} 
\label{eq:disp_1}
\delta = \int_0^{y'_\infty} (1 - u') \, dy'.
\end{equation}
The wall shear stress is defined as the normal gradient of the streamwise velocity at the wall, 
\begin{equation} 
\label{eq:stress_1}
\tau = u'_{y'}(x', \, 0),
\end{equation}
and quantifies the local tangential force per unit area exerted by the fluid on the surface. 
Together, $\delta$ and $\tau$ serve as key diagnostics for assessing the influence of small-scale surface texture and for validating the asymptotic predictions.

\subsection{Asymptotic solution} \label{sec:asymptotic}

\update{While the perturbation procedure follows standard asymptotic methods, the resulting solution plays a central role in the framework by providing explicit scaling relations for how slip modifies displacement thickness and wall shear stress, and by serving as a benchmark against which the numerical solution can be assessed. 
The regime $\lambda \ll 1$ is also relevant for several textured-surface applications discussed in Section~\ref{subsec:SHSs_riblets}. 
}
In this limit, we solve the boundary-layer equations (\ref{eq:middle_1},\,\ref{eq:middle_2},\,\ref{eq:matching_1}) by expanding the velocity fields in a regular perturbation expansion in $\lambda$, obtaining corrections to Blasius’ solution: 
\begin{equation} 
\label{eq:expansion}
u' = u_0 + \lambda u_1/\sqrt{x'} + \lambda^2 u_2/x' + \dots, \quad
v' = v_0 + \lambda v_1/\sqrt{x'} + \lambda^2 v_2/x' + \dots.
\end{equation}
\update{When \(\lambda \ll 1\), we have \(\lambda/\sqrt{x'} \ll 1\) for \(x' = O(1)\). 
This condition fails sufficiently close to the leading edge as \(x' \to 0\), indicating that the slip perturbation expansion developed below is not uniformly valid in \(x'\). 
The numerical solution described in Section~\ref{sec:numerics} remains valid within the boundary-layer model in the regime where slip corrections become \(O(1)\), since it solves the full boundary-layer equations rather than a perturbation expansion. 
As \(\lambda \to 0\), the region where slip effects become comparable to the leading-order solution shrinks towards the leading edge, so the asymptotic expansion remains valid away from the immediate leading-edge region. 
}

\subsubsection{Leading-order solution}

Substituting \eqref{eq:expansion} into (\ref{eq:middle_1},\,\ref{eq:middle_2},\,\ref{eq:matching_1}), the leading-order terms satisfy Blasius' problem for a flat plate. 
Using the similarity variable \(\eta = y' / \sqrt{x'}\), the leading-order velocity can be written as 
\begin{equation} 
\label{eq:blasius_1}
u_0 = f_{\eta}, \quad
v_0 = (\eta f_{\eta} - f)/(2 \sqrt{x'}),
\end{equation}
where \(f = f(\eta)\) satisfies the Blasius ordinary differential equation (ODE) and boundary conditions:
\begin{equation} 
\label{eq:blasius_2}
2 f_{\eta\eta\eta} + f f_{\eta\eta} = 0, \quad f(0) = 0, \quad f_{\eta}(0) = 0, \quad f_{\eta}(\infty) = 1.
\end{equation}
\update{The solution to \eqref{eq:blasius_2} is given in Appendix~\ref{app:Blasius' profile and first-order correction}.} 
Here, $v_0$ is determined from continuity. 
The leading-order displacement thickness \eqref{eq:disp_1} is 
\begin{equation} 
\label{eq:blas_disp}
\delta_0 = \int_0^\infty (1 - f_{\eta}) \, d y' = \sqrt{x'} \int_0^\infty (1 - f_{\eta}) \, d\eta \approx 1.72 \sqrt{x'},
\end{equation}
and the leading-order wall shear stress \eqref{eq:stress_1} is 
\begin{equation} 
\label{eq:blas_stress}
\tau_0 = u_{0y'}(x',\,0) = f_{\eta\eta}(0)\sqrt{x'} \approx 0.33/\sqrt{x'}. 
\end{equation}

\subsubsection{First-order correction}

At \(O(\lambda / \sqrt{x'})\) of \eqref{eq:expansion} in (\ref{eq:middle_1},\,\ref{eq:middle_2},\,\ref{eq:matching_1}), we introduce the fist-order correction 
\begin{equation} 
\label{eq:blasius_4}
u_1 = g_{\eta}, \quad v_1 = (\eta g_{\eta} - g)/(2 \sqrt{x'}),
\end{equation}
where \(g = g(\eta)\) satisfies an ODE, linearised about the Blasius profile $f$, and boundary conditions 
\begin{equation} 
\label{eq:blasius_5}
2 g_{\eta\eta\eta} + f g_{\eta\eta} + f_{\eta\eta} g = 0, \quad g(0) = 0, \quad g_{\eta}(0) = -f_{\eta\eta}(0), \quad g_{\eta}(\infty) \to 0,
\end{equation}
\update{The solution to \eqref{eq:blasius_5} is given in Appendix~\ref{app:Blasius' profile and first-order correction}.} 
The first-order correction redistributes the velocity profile but does not change the wall gradient, as $g_{\eta\eta}(0) = 0$. 

\subsubsection{Second-order solution}

At \(O(\lambda^2 / x')\) of \eqref{eq:expansion} in (\ref{eq:middle_1},\,\ref{eq:middle_2},\,\ref{eq:matching_1}), we introduce the second-order correction 
\begin{equation} 
\label{eq:blasius_6}
u_2 = h_{\eta}, \quad v_2 = (\eta h_{\eta} - h)/(2 \sqrt{x'}),
\end{equation}
where \(h = h(\eta)\) satisfies an ODE, linearised about the leading-order Blasius solution $f$ and first-order correction $g$, and boundary conditions 
\begin{equation} 
\label{eq:blasius_7}
2 h_{\eta\eta\eta} + f h_{\eta\eta} + f_{\eta\eta} h + g g_{\eta} = 0, \quad h(0) = 0, \quad h_{\eta}(0) = -g_{\eta\eta}(0), \quad h_{\eta}(\infty) \to 0.
\end{equation}

\subsubsection{Displacement thickness and wall shear stress}

The corrected displacement thickness \eqref{eq:disp_1} is 
\begin{equation} 
\label{eq:delta_corrected}
\delta = \int_0^\infty (1 - f_{\eta} + \lambda g_{\eta}/\sqrt{x'} +\dots) \sqrt{x'} \, d\eta = \sqrt{x'} (1.72 - \lambda /\sqrt{x'} + \dots),
\end{equation}
and the corrected wall shear stress \eqref{eq:stress_1} is 
\begin{multline} 
\label{eq:shear_corrected}
\tau = u'_{y'}(x',\,0) = (f_{\eta\eta}(0) + \lambda g_{\eta\eta}(0)/\sqrt{x'} + \lambda^2 h_{\eta\eta}(0)/x' +\dots)/\sqrt{x'} \\ = (0.33 - 0.15 \lambda^2/x' + \dots)/\sqrt{x'},
\end{multline}
since $g_{\eta\eta}(0)=0$. 
Physically, the first-order correction alters the velocity profile without changing the wall shear; the first non-zero contribution to the wall shear arises at second order. 
These asymptotic results for \(\delta\) and \(\tau\) quantify how small-scale surface slip alters key boundary-layer properties such as drag and thickness, and provide a benchmark for validating the numerical simulations discussed in Section~\ref{sec:numerics}. 

\section{Stability} \label{sec:stability}

\update{We now examine the linear stability of the slip-modified boundary-layer flows derived in Section~\ref{sec:model}. 
Although the base flow is obtained from a steady boundary-layer formulation with a slip boundary condition, the stability analysis is performed by linearising the unsteady Navier--Stokes equations about this base state. 
This procedure is standard in boundary-layer stability theory, where asymptotically derived base flows are used within the full equations and the problem is formulated locally at a fixed streamwise position \(x=x_0\), invoking a locally parallel-flow approximation~\cite{schmid2012stability}. 
This approximation is valid when the streamwise variation of the base flow occurs on length scales large compared with the disturbance wavelength, which holds in the present high-Reynolds-number regime. 
Accordingly, the present study performs a local stability analysis of the boundary-layer region; a complementary global stability analysis of the inner region, which may exhibit additional or stronger instability mechanisms associated with the viscous roughness region, is beyond the scope of this paper. 
}

\update{Let $u'=u'(y)$ denote the steady streamwise velocity component of the solution of the slip boundary-layer equations~(\ref{eq:middle_1}, \ref{eq:middle_1b}, \ref{eq:middle_2}, \ref{eq:matching_1}), evaluated at \(x=x_0\). 
We consider perturbations of amplitude $\Delta$ superposed on this base flow of normal-mode form
\begin{align}
u(x,\,y,\,t) &= u'(y) + \Delta \hat{u}(y)\,\mathrm{e}^{i(\alpha x - \omega t)} + \text{c.c.}, \label{eq:pert_u}\\
v(x,\,y,\,t) &= \Delta \hat{v}(y)\,\mathrm{e}^{i(\alpha x - \omega t)} + \text{c.c.}, \label{eq:pert_v}\\
p(x,\,y,\,t) &= \Delta \hat{p}(y)\,\mathrm{e}^{i(\alpha x - \omega t)} + \text{c.c.}, \label{eq:pert_p}
\end{align}
where \(\alpha\) is the streamwise wavenumber, \(\omega\) is the complex frequency and c.c. denotes the complex conjugate. 
Substituting~\eqref{eq:pert_u}--\eqref{eq:pert_p} into the non-dimensional Navier--Stokes equations~\eqref{eq:nondim_1}--\eqref{eq:nondim_1c} and linearising in $\Delta$ yields the following system for the perturbation fields:
\begin{align}
i\alpha \hat{u} + \hat{v}_y &= 0, \label{eq:lin_cont}\\
-i\omega \hat{u} + i\alpha u' \hat{u} + \hat{v} u'_y  &=
- i \alpha \hat{p} + \frac{1}{{Re}}\left(\hat{u}_{yy} - \alpha^2 \hat{u}\right), \label{eq:lin_u}\\
-i\omega \hat{v} + i\alpha u' \hat{v} &=
-\hat{p}_y + \frac{1}{{Re}}\left(\hat{v}_{yy} - \alpha^2 \hat{v}\right). \label{eq:lin_v}
\end{align} 
Eliminating $\hat{u}$ using continuity~\eqref{eq:lin_cont} and differentiating~\eqref{eq:lin_u} with respect to $y$ to remove the pressure term reduces \eqref{eq:lin_cont}–\eqref{eq:lin_v} to the Orr--Sommerfeld equation for the normal-velocity perturbation,
\begin{equation}
\left(\mathrm{d}_{yy}-\alpha^2\right)^2 \hat{v}
- i\alpha {Re}\left[
u'\left(\mathrm{d}_{yy}-\alpha^2\right)\hat{v}
- u'_{yy} \hat{v}
\right]
= -i\omega {Re} \left(\mathrm{d}_{yy}-\alpha^2\right)\hat{v}.
\label{eq:OS_v}
\end{equation} 
The Orr--Sommerfeld equation \eqref{eq:OS_v} is subject to the slip boundary condition and the requirement of no penetration at the surface in the boundary-layer region \eqref{eq:full_slippy}. 
Using continuity~\eqref{eq:lin_cont} to eliminate \(\hat{u}\),  \eqref{eq:full_slippy} reduces to
\begin{equation}
\hat{v}_y(0)= \lambda \hat{v}_{yy}(0) / {Re}^{1/2}, \qquad \hat{v}(0)=0.\vspace{.1cm}
\label{eq:OS_bc_wall}
\end{equation} 
As \(y\to\infty\), disturbances decay into the free stream~\eqref{eq:middle_2}, giving
\begin{equation}
\hat{v}_y \to 0, \qquad \hat{v} \to 0.
\label{eq:OS_bc_inf}
\end{equation} 
Equations~\eqref{eq:OS_v}--\eqref{eq:OS_bc_inf} define an eigenvalue problem for \(\omega\), which generalises the classical Orr--Sommerfeld formulation for the Blasius boundary layer to include the effects of wall slip. 
}

\subsection{Numerical solution}
\label{sec:stability_numerics}

\update{
The Orr--Sommerfeld eigenvalue problem subject to the slip boundary conditions~\eqref{eq:OS_v}--\eqref{eq:OS_bc_inf} is also solved numerically using a Chebyshev spectral collocation methods in the normal direction, similar to the base flow in Section~\ref{sec:numerics}. 
The base flow $u'$ entering~\eqref{eq:OS_v} is obtained from the boundary-layer computation described in Section~\ref{sec:numerics} and is treated as locally parallel, $v'=0$, at a fixed streamwise location, $x=x_0$. 
The normal domain $y\in[0,\,y_\infty]$ is mapped linearly to the standard Chebyshev domain $\xi\in[-1,\,1]$. 
Once again, the truncation height $y_\infty$ is chosen sufficiently large to ensure that both the base flow and perturbations have decayed to their free-stream values, ensuring that~\eqref{eq:OS_bc_inf} is enforced. 
Spectral differentiation matrices $D_\xi$ and $D_\xi^2$ are constructed following standard Chebyshev-collocation procedures~\cite{trefethen2000spectral} and scaled to the physical domain. 
Let $\hat{\mathbf{v}}\in\mathbb{C}^{N_\xi+1}$ denote the discrete solution vector of values of the normal-velocity eigenvectors at the collocation points. 
Discretising the Orr--Sommerfeld equations~\eqref{eq:OS_v} yields a generalised eigenvalue problem of the form:
\begin{equation}
\mathbf{C}\hat{\mathbf{v}} = \omega \mathbf{D}\hat{\mathbf{v}},
\label{eq:OS_evp}
\end{equation}
where sub-matrices $\mathbf{C}$ and $\mathbf{D}\in\mathbb{C}^{(N_\xi+1)\times(N_\xi+1)}$ encode the differential operators depending on the base flow and perturbation waveform. 
Boundary conditions are enforced by replacing the appropriate rows of $\mathbf{C}$ and $\mathbf{D}$. 
At the wall ($y=0$), the no-penetration condition and the linearised slip condition~\eqref{eq:OS_bc_wall} are imposed, $\hat{v}=0$ and $\hat{v}_y = \lambda \hat{v}_{yy}/{Re}^{1/2}$, while at the outer boundary ($y=y_\infty$) the decay conditions~\eqref{eq:OS_bc_inf} are enforced, $\hat{v}=0$ and $\hat{v}_y=0$. 
For a given $Re$, $\lambda$ and $\alpha$, the generalised eigenvalue problem~\eqref{eq:OS_evp} is solved using standard eigenvalue solvers (eig.m) in MATLAB. 
The resulting spectrum $\omega=\omega_r+i\omega_i$ determines the temporal stability of the flow, with eigenmodes satisfying $\omega_i>0$ corresponding to linear instability. 
Convergence of the eigenvalues and eigenvectors was verified by increasing both the number of collocation points $N_\xi$ and the domain height $y_\infty$ until changes in the most unstable eigenvalues were negligible. 
The stability results presented below use wall-normal resolution $N_\xi = 100$ and domain truncation $y_\infty = 30$ as employed for the base-flow computations. 
}

\section{Results} \label{sec:results}

\begin{figure*}[t!]
    \centering
    (a) \hfill (b) \hfill \hfill \hfill \\
    \includegraphics[width=0.5\linewidth]{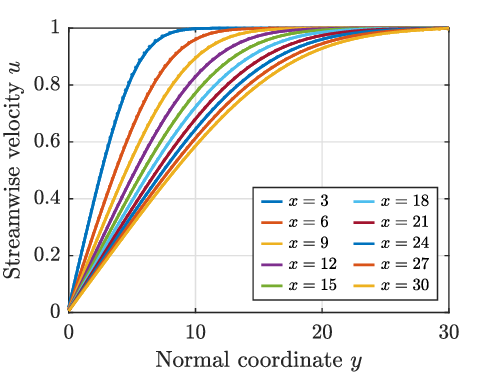}\includegraphics[width=0.5\linewidth]{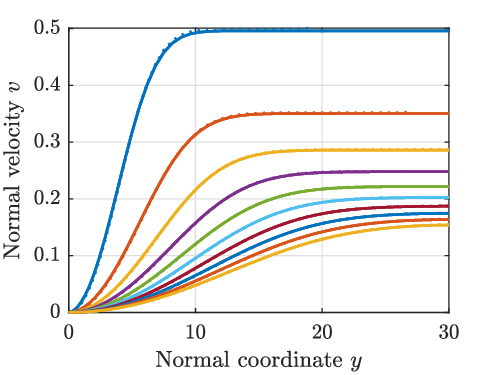} \\
    (c) \hfill (d) \hfill \hfill \hfill \\
    \includegraphics[width=0.5\linewidth]{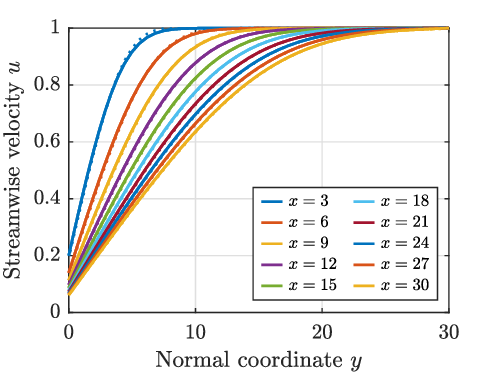}\includegraphics[width=0.5\linewidth]{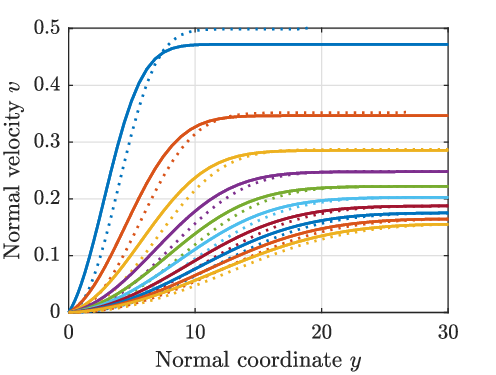} \\
    \caption{Streamwise velocity $u$ and normal velocity $v$ for (a--b) $\lambda = 0.1$ and (c--d) $\lambda = 1$. 
    Dashed lines indicate asymptotic solutions using (\ref{eq:expansion}, \ref{eq:blasius_1}, \ref{eq:blasius_4}); solid lines show numerical results using \eqref{eq:numerical_sol}. 
    The plots illustrate how slip modifies the boundary-layer development, with larger deviations near the leading edge for larger $\lambda$. 
    }
    \label{fig:2}
\end{figure*}

\subsection{Streamwise and normal velocity field} \label{subsec:Streamwise and normal velocity field}


In this section, we present the evolution of the streamwise velocity $u'(x', \,y')$ and the normal velocity $v'(x', \,y')$ over a small-scale textured surface subject to the slip boundary condition \eqref{eq:matching_1}. 
We compare asymptotic results (Section~\ref{sec:asymptotic}) and numerical simulations (Section~\ref{sec:numerics}) to quantify how slip modifies boundary-layer development. 
At this stage, the slip length $\lambda = \epsilon \beta \text{Re}^{1/2}$ is treated as a parameter representing the influence of small-scale surface texture, without reference to a specific geometry. 

Figure~\ref{fig:2} illustrates the streamwise and normal velocity fields for two representative slip lengths, $\lambda = 0.1$ and $\lambda = 1$, over the streamwise domain $x' \in [0, \,30]$ and the normal domain $y' \in [0, \,30]$. 
The numerical results are plotted using solid lines, and the asymptotic predictions using dashed lines. 
For $\lambda = 0.1$, the slip length remains small relative to the boundary-layer thickness ($\sim \sqrt{x'}$), so that $\lambda/\sqrt{x'} \ll 1$, corresponding to a small-slip regime. 
In this regime, the small-scaled textured surface slightly reduces the wall shear without significantly altering the boundary-layer structure. 
The numerical and asymptotic solutions remain in agreement throughout the domain, including near the leading edge. 
For $\lambda = 1$, the assumption $\lambda/\sqrt{x'} \ll 1$ breaks down near the leading edge; for example, at $x' = 3$, $\lambda/\sqrt{3} \approx 0.577$, which is no longer small. 
This results in small deviations between numerical and asymptotic solutions near the leading edge, as slip modifies the velocity field. 
Downstream, as the boundary layer thickens and $\lambda/\sqrt{x'}$ decreases, agreement between numerical and asymptotic solutions improves. 

At the wall ($y' = 0$), the non-zero slip velocity $u'(x',\, 0) > 0$ reduces the velocity gradient and the wall shear stress \eqref{eq:stress_1}. 
Away from the wall, $u'$ increases monotonically towards the free-stream value $u' \rightarrow 1$ as $y' \rightarrow \infty$, while the boundary-layer thickness grows downstream $\sim \sqrt{x'}$. 
The normal velocity $v'$ vanishes at the wall due to the no-penetration condition and increases with distance from the wall within the boundary layer. 
For fixed $x'$, $v'$ reaches a maximum at a height scaling as $\sqrt{x'}$, while the maximum decreases as $1/\sqrt{x'}$, consistent with similarity arguments and the leading-order asymptotic solution (Section~\ref{sec:asymptotic}). 
Increasing $\lambda$ enhances the slip velocity at the wall, further reducing the near-wall velocity gradient and wall shear. 
This accelerates the approach of $u'$ to its free-stream value and modifies the distribution of $v'$. 
In particular, for larger $\lambda$, $u'$ reaches $1$ for smaller $y'$, while $v'$ attains larger values far away from the wall ($y' \gg 1$). 
Near the leading edge, the asymptotic approximation is less accurate for larger $\lambda$,  leading to larger differences between numerical and asymptotic results for $v'$ in particular. 
Further downstream, both $u'$ and $v'$ converge toward the asymptotic predictions for all $\lambda$. 

\subsection{Displacement thickness and wall shear stress} \label{subsec:Displacement thickness and wall shear stress}

\begin{figure*}[t!]
    \centering
    (a) \hfill (b) \hfill \hfill \hfill \\
    \includegraphics[width=0.5\linewidth]{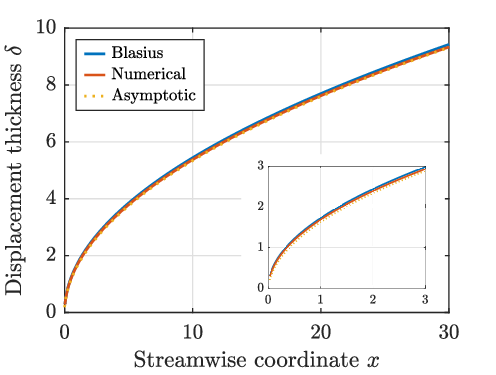}\includegraphics[width=0.5\linewidth]{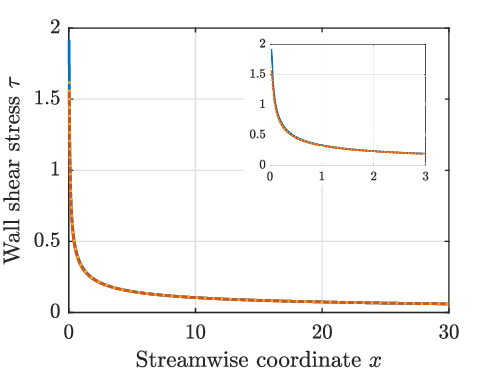} \\
    (c) \hfill (d) \hfill \hfill \hfill \\
    \includegraphics[width=0.5\linewidth]{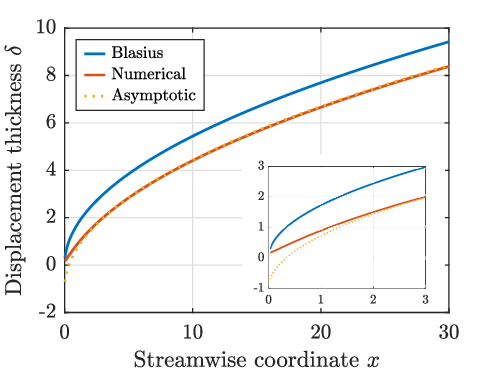}\includegraphics[width=0.5\linewidth]{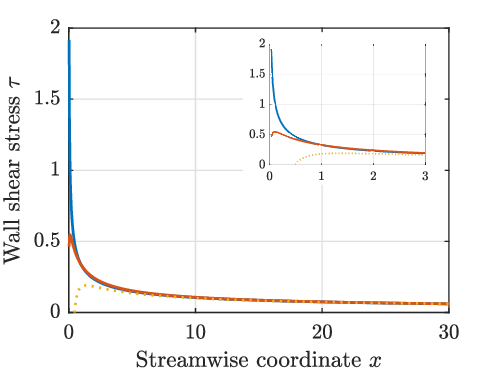} \\
    \caption{Displacement thickness $\delta$ and wall shear stress $\tau$ for (a--b) $\lambda = 0.1$ and (c--d) $\lambda = 1$. 
    Dotted lines depict asymptotic solutions using (\ref{eq:delta_corrected}, \ref{eq:shear_corrected}); solid red lines show numerical results obtained from (\ref{eq:numerical_sol}, \ref{eq:disp_1}, \ref{eq:stress_1}); solid blue lines indicate the Blasius solution (\ref{eq:blas_disp}, \ref{eq:blas_stress}). 
    The plots illustrate how slip modifies boundary-layer thickness and wall shear, with larger deviations near the leading edge for higher $\lambda$. 
    }
    \label{fig:3}
\end{figure*}

We now examine two key quantities that characterise the boundary layer: the displacement thickness $\delta(x')$ \eqref{eq:disp_1} and the wall shear stress $\tau(x')$ \eqref{eq:stress_1}. 
These quantities quantify the momentum deficit and surface drag, providing integral measures of the boundary layer’s response to slip induced by small-scale surface texture. 
Figure \ref{fig:3} shows the streamwise variation of $\delta$ and $\tau$ for slip lengths $\lambda = 0.1$ and $\lambda = 1$ over the domain $x' \in [0, \,30]$. 

The displacement thickness represents the equivalent loss of mass flux due to the presence of the boundary layer. 
In Blasius' theory, $\delta \sim \sqrt{x'}$ for $x' \gg 1$, a consequence of the self-similarity of the velocity field~\cite{schlichting2017boundary}. 
For small slip (e.g., $\lambda = 0.1$ in Figure~\ref{fig:3}(a)), the numerical results closely match both the Blasius solution and the asymptotic solution (Section \ref{sec:asymptotic}), which remains valid across the domain in the small-slip regime. 
The small-slip displacement thickness \eqref{eq:delta_corrected} takes the form $\delta = \delta_0 + \lambda \, \delta_1/ \sqrt{x'} + \dots$, where $\delta_0 \sim \sqrt{x'}$ is the Blasius contribution and $\lambda \, \delta_1/ \sqrt{x'}$ is an $O(1)$ correction arising from the slip boundary condition. 
This correction leads to a uniform reduction in $\delta$ across $x'$, indicating that the slip lessens the velocity deficit in the near-wall region. 
For larger slip (e.g., $\lambda = 1$ in Figure~\ref{fig:3}(c)), the numerical and asymptotic results diverge near the leading edge, reflecting the breakdown of the $\lambda/\sqrt{x'}\ll 1$ assumption in this region. 
In this region, the boundary layer is thin and highly sensitive to slip, resulting in behaviour not captured by the regular expansion \eqref{eq:delta_corrected}. 
Further downstream, the numerical displacement thickness approaches the Blasius value, demonstrating that even moderate slip primarily affects the near-wall region during the early stages of boundary-layer development. 

In Blasius' theory, the wall shear stress scales as $\tau \sim 1/\sqrt{x'}$~\cite{schlichting2017boundary}. 
For flows with slip, the wall shear stress is modified due to changes in near-wall velocity gradients. 
The asymptotic prediction for $\tau$, obtained from the corrected similarity solution~\eqref{eq:shear_corrected}, exhibits the expected $1/\sqrt{x'}$ decay for large $x'$. 
A higher-order $O(\lambda/x'^{3/2})$ correction becomes significant for small $x'$, since the $O(\lambda/x')$ contribution turns out to be zero in the solution of \eqref{eq:blasius_5}. 
For small slip (e.g., $\lambda = 0.1$, Figure~\ref{fig:3}(b)), numerical and asymptotic solutions coincide across the domain. 
For larger slip (e.g., $\lambda = 1$, Figure~\ref{fig:3}(d)), the numerical wall shear stress deviates from the asymptotic prediction near the leading edge, where $O(\lambda/\sqrt{x'})$ terms are no longer small. 
In this region, both numerical and asymptotic solutions exhibit a non-monotonic rise before converging to the expected $1/\sqrt{x'}$ decay. 
This initial increase likely arises from the interaction between the high slip velocity at the wall and steep near-wall gradients, a behaviour absent in Blasius' solution. 
For large $x'$, all three solutions (numerical, asymptotic and Blasius) converge, as the influence of slip becomes negligible (Figure~\ref{fig:3}(b, d)). 
Increasing $\lambda$ reduces the wall shear stress, meaning that slip lowers the viscous drag $D = \int_{x'=0}^L \tau \,\text{d}x'$ over a surface of length $L$~\cite{schlichting2017boundary}, consistent with the role of slip in drag-reduction technologies. 

\subsection{Stability}

\begin{figure*}[t!]
    \centering
    (a) \hfill (b) \hfill \hfill \hfill \\
    \includegraphics[width=0.49\linewidth]{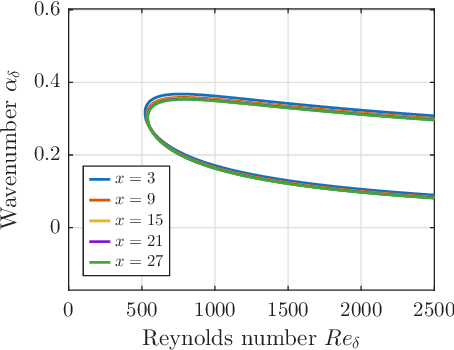}\includegraphics[width=0.5\linewidth]{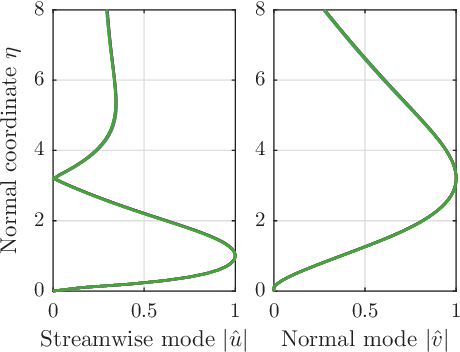} \\
    (c) \hfill (d) \hfill \hfill \hfill \\
    \includegraphics[width=0.49\linewidth]{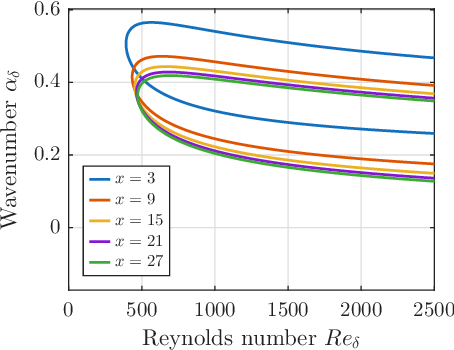}\includegraphics[width=0.5\linewidth]{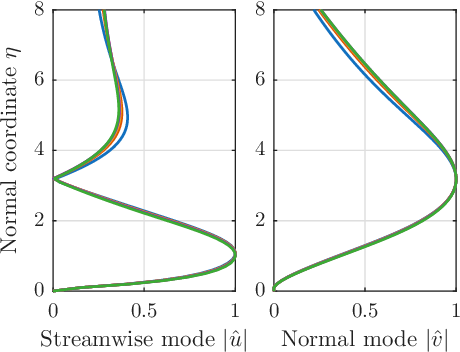} \\
    \caption{\update{Neutral curves $\text{Im}(\omega)=0$ and the most unstable velocity eigenmodes $|\hat{u}|$ and $|\hat{v}|$ for (a--b) $\lambda = 0.1$ and (c--d) $\lambda = 1$ at $x=3$, 9, 15, 21, 27. 
    The plots illustrate how slip modifies the boundary-layer stability, with the flow becoming more unstable near the leading edge for larger $\lambda$. 
    As $\lambda \rightarrow 0$, we recover the critical wavenumber and Reynolds numbers given in Table 3.1 of Schmit and Henningson~\cite{schmid2012stability}, $\alpha_{\delta} \approx 0.3$ and ${Re}_{\delta} \approx 520$, where our $\alpha$ and ${Re}$ are multiplied by a the displacement thickness 1.72 to account for the different definitions. 
    }
    }
    \label{fig:3_new}
\end{figure*}

\update{
We next examine the linear stability of the slip-modified boundary layer by computing neutral curves from the Orr--Sommerfeld system~(\ref{eq:OS_v})--(\ref{eq:OS_bc_inf}), identifying conditions for which the imaginary part of the wave frequency is zero. 
Results are presented for two representative slip lengths, $\lambda=0.1$ and $\lambda=1$, at streamwise locations, $x=3$, 9, 15, 21 and 27, as shown in Figure~\ref{fig:3_new}.  
For sufficiently large streamwise distances, the neutral curves closely resemble those obtained for the classical Blasius boundary layer~\cite{schmid2012stability}, indicating that slip effects become progressively weaker downstream. 
}

\update{For small slip ($\lambda=0.1$, Figure~\ref{fig:3_new}(a--b)), the neutral curves and critical parameters remain close to the no-slip case across the domain, consistent with slip acting as a small perturbation to the classical Tollmien--Schlichting instability. 
As the slip length increases ($\lambda=1$, Figure~\ref{fig:3_new}(c--d)), the unstable wavenumber band shifts to higher values and the critical Reynolds number decreases, indicating an enhanced susceptibility of the laminar boundary layer to instability. 
The corresponding eigenmodes retain the characteristic Tollmien--Schlichting structure, with velocity perturbations concentrated within the boundary layer, although their wall-normal distribution varies modestly with streamwise position for $\lambda = 1$. 
Overall, these results demonstrate that sufficiently strong slip can locally modify the stability characteristics of laminar boundary layers within the present homogenised framework. 
}

\subsection{Slip lengths for SHSs and riblets} \label{subsec:SHSs_riblets}

\begin{figure*}[t!]
    \centering
    (a) \hfill (b) \hfill \hfill \hfill \\
    \includegraphics[width=0.5\linewidth]{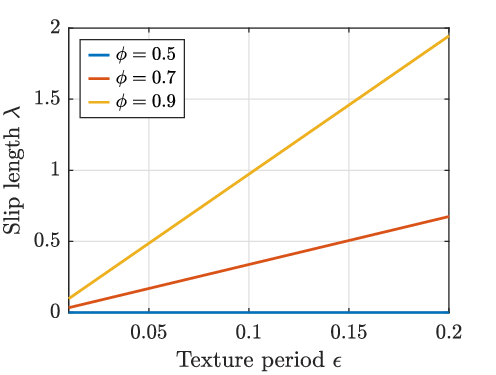}\includegraphics[width=0.5\linewidth]{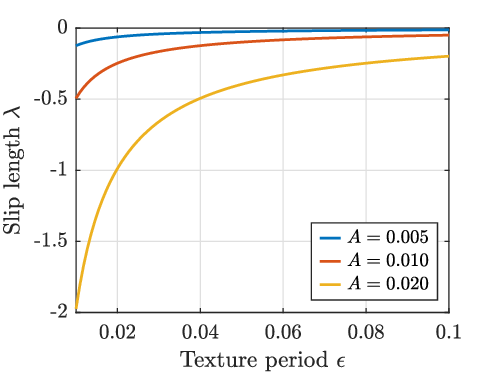} \\
    \caption{Slip length $\lambda$ for (a) transverse SHSs as a function of gas fraction $\phi$ and half-period $\epsilon$ and (b) transverse riblets as a function of texture amplitude $A$ and half-period $\epsilon$, where $\text{Re} = 100$.  
    The plots show how geometry and texture parameters influence slip behaviour in laminar boundary layers. 
    }
    \label{fig:4}
\end{figure*}

To provide practical context, we consider the slip lengths for two common textured surfaces: SHSs and riblets. 
Figure \ref{fig:4}(a) shows how the slip length $\lambda$  varies with geometric parameters of SHSs, including the texture half-period $\epsilon$ and gas fraction $\phi$. 
For SHSs with transverse ridges that trap shear-free air pockets beneath the liquid--gas interface, the slip length is given by \eqref{eq:philip}. 
This expression assumes a periodic array of ridges with alternating no-slip and shear-free regions and a flat liquid–gas interface. 
It is derived by solving the Stokes flow problem over a representative periodic cell using conformal mapping and averaging the tangential velocity at the interface. 
The slip length increases with $\phi$, $\epsilon$ and $\text{Re}$, reflecting the larger proportion of shear-free area, the expanded period size and the thinner boundary layer. 
In contrast, transverse riblets increase drag in laminar flow due to the additional surface area and flow obstruction. 
The negative slip length shown in Figure \ref{fig:4}(b) corresponds to a reduction in effective wall velocity relative to the no-slip case, resulting in higher wall shear. 
For riblets of sinusoidal profile with small amplitude $A \ll \epsilon$, the slip length is given asymptotically by \eqref{eq:luchini}. 
This expression is obtained by solving the inner Stokes problem over the sinusoidal wall geometry and matching it to the outer boundary-layer solution. 
The negative sign indicates that the near-wall velocity is lower than in the no-slip case, leading to a higher velocity gradient and increased wall shear (drag). 

\update{Figure~\ref{fig:4} shows that the values of \(\lambda\) considered in Sections~\ref{subsec:Streamwise and normal velocity field}--\ref{subsec:Displacement thickness and wall shear stress} arise from geometric and flow parameters representative of SHS and riblet applications. 
Table~\ref{tab:shs_riblet_ranges} summarises representative values of $\epsilon$, $\phi$, $A$ and $\text{Re}$ for SHSs and riblets in micro-fluidic devices, turbo-machinery blades, aircraft and marine applications~\cite{tomlinson2023laminar, tomlinson2023model, ananth2025performance, yang2021numerical}. 
These parameter ranges indicate that the assumptions $\epsilon \ll \text{Re}^{-1/2}\ll 1$ for SHSs and $A \ll \epsilon \ll \text{Re}^{-1/2}\ll 1$ for riblets are reasonably satisfied in many practical applications, supporting the use of the present three-region asymptotic framework. 
In some applications in Table~\ref{tab:shs_riblet_ranges}, these scale separations are only weakly satisfied, and the model should therefore be interpreted as providing qualitative trends rather than a quantitative description. 
In such regimes, quantitative assessment requires comparison with texture-resolving numerical simulations or experiments for specific surfaces and flow conditions. 
For low- and moderate-Reynolds-number flows (${Re}\lesssim10^{5}$), such as channel flows and microfluidic devices, the boundary layer is laminar and the present theory applies without modification. 
In higher-Reynolds-number applications (${Re}\gtrsim10^{5}$), including aircraft and marine surfaces, the boundary layer is predominantly turbulent; however, it remains laminar near the leading edge prior to transition. 
The present framework is therefore applicable to this upstream laminar region, while extension of the approach to turbulent boundary layers lies beyond the scope of the present work. 
}

\begin{table*}[t!]
\centering
\caption{Representative parameter ranges for SHSs and riblets across different applications~\cite{tomlinson2023laminar, tomlinson2023model, ananth2025performance, yang2021numerical}. 
$\epsilon = \hat{P}/\hat{L}$ denotes the texture half-period, $\phi$ is the gas fraction of the SHS, $A = \hat{A}/\hat{L}$ is the amplitude of the riblet and \(\text{Re}= \hat{U}\hat{L}/\hat{\nu}\) denotes the Reynolds number. 
}
\small
\vspace{.5cm}
\begin{tabular}{llccc}
\toprule \addlinespace
Type & Application & \(\epsilon\) & \(\phi\) or \(A\) & \(\text{Re}\) \\ \addlinespace
\midrule \addlinespace
SHS
& Channels and microfluidics        & \(0.01 - 0.05\) & \(\phi = 0.5 - 0.9\)   & \(10^{-2} - 10^{2}\)       \\ \addlinespace
& Marine vessels     & \(0.001 - 0.005\) & \(\phi = 0.3 - 0.7\)   & \(10^{7} - 10^{8}\)        \\
\midrule \addlinespace
Riblet 
& Turbo-machinery blades         & \(0.01 - 0.05\) & \(A = 0.002 - 0.01\)   & \(10^{4} - 10^{6}\)        \\ \addlinespace
& Aircraft surfaces             & \(0.001 - 0.01\)  & \(A = 0.005 - 0.015\)  & \(10^{6} - 10^{8}\)        \\ \addlinespace
\bottomrule
\end{tabular}
\label{tab:shs_riblet_ranges}
\end{table*}

\section{Discussion}
\label{sec:discussion}

\update{We have developed a framework for steady, 2D laminar boundary layers over small-scale textured surfaces, in which the surface geometry is incorporated through a slip length.} 
This formulation combines matched asymptotic expansions with numerical computation, enabling the flow to be resolved without directly simulating the small-scale texture. 
A main contribution of this work is a slip-corrected boundary-layer model, in which the Prandtl boundary-layer equations are solved with a slip boundary condition of the form $u = \lambda u_y$ imposed at the wall. 
This slip boundary condition captures the influence of small-scale surface texture through a macroscopic parameter, enabling efficient prediction of drag and boundary-layer quantities. 
\update{This formulation provides a framework in which a variety of textured surfaces can be compared consistently through their slip length.} 
The slip length is obtained by homogenising the inner viscous layer and is valid under the assumption that the texture half-period $\hat{P}$ is small compared to the \update{reference length} $\hat{L}$, i.e., $\epsilon = \hat{P}/\hat{L} \ll 1$. 
Under this scale separation, the small-scale texture perturbs the boundary-layer flow only through its averaged effect. 

We have derived an asymptotic solution for the velocity field valid when the slip length is small compared with the boundary-layer thickness, i.e., $\lambda/\sqrt{x} \ll 1$, giving a perturbation to the Blasius solution with the correction arising from wall slip. 
This condition arises by comparing the slip velocity, of order $\lambda u_y$, with the streamwise velocity $u$ in the outer region. 
When $\lambda / \sqrt{x}$ is small, slip acts as a weak perturbation and the similarity structure is preserved. 
\update{In the small-slip regime, the asymptotic solution provides accurate and inexpensive closed-form estimates of displacement thickness \eqref{eq:delta_corrected} and wall shear stress \eqref{eq:shear_corrected}, and defines an analytical limit against which the numerical scheme can be validated (and vice versa). 
}
To extend the model to cases where $\lambda / \sqrt{x} = O(1)$, we developed a numerical scheme that employs Chebyshev collocation in the wall-normal direction and implicit marching in the streamwise direction, which provides stability and high accuracy in resolving steep velocity gradients near the wall. 
Numerical results (Figure~\ref{fig:2}--\ref{fig:3}) confirm the asymptotic predictions in the regime $\lambda/\sqrt{x} \ll 1$ and extend them to larger slip lengths. 
Both approaches show that the slip-corrected velocity field, wall shear stress and displacement thickness deviate most strongly from the Blasius solution near the leading edge, where the boundary layer is thin and velocity gradients are steep, before converging downstream as $\lambda / \sqrt{x}$ decreases. 
\update{
We have also examined the linear stability of the slip-modified boundary layers using a local, parallel-flow approximation. 
The stability analysis shows that wall slip influences disturbance growth near the leading edge, where the slip length relative to the boundary-layer thickness is largest. 
As the slip length increases, the neutral stability boundary shifts to lower Reynolds numbers and higher amplification rates are observed, indicating that slip enhances instability in this region. 
}

The present analysis is restricted to steady, 2D, laminar boundary layers. 
While appropriate for many applications, the current framework does not account for unsteady or three-dimensional (3D) effects~\cite{tomlinson2024unsteady}, which could be incorporated in future extensions. 
\update{Three-dimensionality is discussed further in Appendix~\ref{app:3d}.} 
The assumption of a high Reynolds number with a thin boundary layer justifies the use of Prandtl boundary-layer theory, but excludes low-Reynolds-number regimes where viscous effects dominate the entire flow domain. 
In practice, textured surfaces may also be random, aperiodic or hierarchical~\cite{sbragaglia2007effective, priezjev2011molecular, cottin2012scaling}. 
Although the present analysis focuses on periodic textures, homogenisation methods can be extended to disordered geometries using stochastic or statistical approaches~\cite{sbragaglia2007effective, ponomarev2003surface}. 
Another idealisation concerns the flatness of the liquid--gas interface on SHSs, which underlies Philip’s analytical slip-length expression~\cite{philip1972flows}. 
In experiments, meniscus curvature and deformation are often significant, leading to deviations from the theoretical slip predictions~\cite{joseph2006slippage, ybert2007achieving}. 
Similarly, the asymptotic model for riblets, expressed as $\beta = -\pi^2 A^2 / (4 \epsilon^2)$, assumes small amplitude ($A \ll \epsilon$) and low curvature~\cite{luchini1991resistance, bechert1997experiments}. 
When this assumption is violated, for example at larger riblet amplitudes, the effective slip may differ from the asymptotic prediction 
\update{
The stability analysis is also restricted to 2D disturbances and a local parallel-flow assumption, consistent with classical Blasius stability theory~\cite{schmid2012stability}.
A global stability analysis is being pursued as future work. 
}

Despite these limitations, the framework provides a foundation for more general problems. 
It can be extended to transitional or turbulent boundary layers~\cite{tomlinson2022linear, tomlinson2023model}, where slip modifies near-wall turbulence structures, or to time-dependent flows~\cite{tomlinson2024unsteady}, such as pulsatile or start-up boundary layers. 
The slip boundary condition may also be generalised to a tensorial form to represent anisotropic textures such as grooves or riblets that are not aligned with the main flow direction~\cite{bazant2008tensorial}. 
By coupling the present boundary-layer formulation to macroscopic flow solvers, it is possible to incorporate microscale surface effects into large-scale simulations without explicitly resolving the surface geometry~\cite{garcia2011drag, bottaro2019flow}. 
Representative applications include SHSs and LISs, where analytical slip-length expressions are available for simple geometries~\cite{joseph2006slippage, philip1972flows, lauga2003effective, ybert2007achieving, wong2011bioinspired, hardt2022flow, sundin2022slip, van2017substantial} and riblets, whose small-amplitude slip effects can be derived asymptotically~\cite{bazant2008tensorial, feuillebois2009effective, zhou2013effective, sharma2020influence}.  
Porous surfaces, where slip follows from Brinkman theory;~\cite{beavers1967boundary, ochoa1995momentum, nield2009beavers}; and rough or randomly textured surfaces, where homogenised or empirical slip lengths are used~\cite{sbragaglia2007effective, ponomarev2003surface, priezjev2011molecular, cottin2012scaling}, can also be accommodated. 
Compliant or deformable substrates, and those subject to Marangoni stresses~\cite{tomlinson2023laminar}, may also be accommodated, although only approximate or experimental slip lengths are typically available~\cite{skotheim2004soft, skotheim2005soft, pandey2016lubrication}. 
\update{The ability to represent these diverse surface textures within a single laminar boundary-layer framework highlights the versatility and generality of the present approach.} 

\section{Conclusions}

This work establishes a framework for modelling laminar boundary-layer flows over small-scale textured surfaces, e.g., SHSs, riblets, LISs, porous compliant or deformable surface, incorporating slip lengths derived from homogenisation. 
\update{The framework captures both the steady boundary-layer structure and the linear stability characteristics of slip-modified flows within a computationally-efficient model.}  
By connecting small-scale surface geometry to large-scale flow dynamics, the model enables accurate prediction of wall shear stress and drag without explicitly resolving the texture. 
\update{As a result, the framework offers simple predictions and a practical route for analysing, designing and optimising small-scale textured surfaces across a range of applications, including microfluidic devices, turbo-machinery blades and marine transport operating in laminar flow regimes.}  

\appendix

\section{Blasius' solution and first-order correction} \label{app:Blasius' profile and first-order correction}

\begin{figure*}
    \centering
    (a) \hfill (b) \hfill \hfill \hfill \\
    \includegraphics[width=0.5\linewidth]{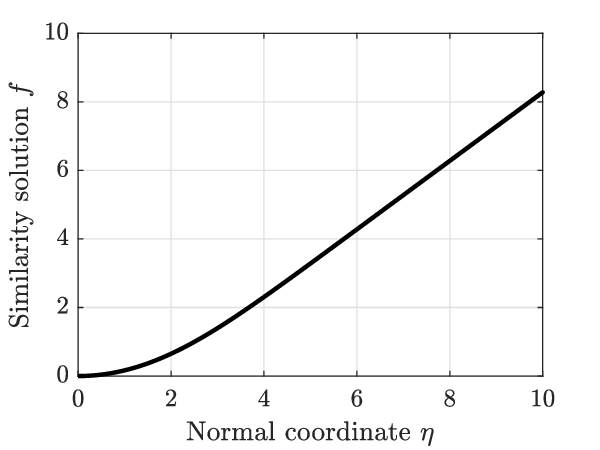}\includegraphics[width=0.5\linewidth]{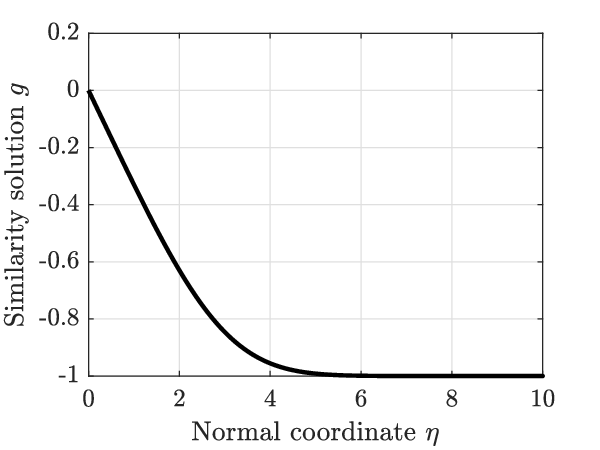}
    \caption{\update{Leading-order and first-order similarity solutions for the velocity profile in the presence of slip. 
    (a) The Blasius profile $f$, satisfying the no-slip boundary condition (\ref{eq:blasius_1}--\ref{eq:blasius_2}). 
    (b) The first-order correction $g$, arising at $O(\lambda/\sqrt{x'})$ from the slip boundary condition at the wall (\ref{eq:blasius_4}--\ref{eq:blasius_5}). 
    }
    }
    \label{fig:sim_sol}
\end{figure*}

\update{Numerical solutions for \(f\) and \(g\) are shown in Figure~\ref{fig:sim_sol}(a) and (b). 
These similarity profiles give the leading-order and first-order terms in the asymptotic boundary-layer expansion \eqref{eq:expansion} for the streamwise and normal velocity field, which leads directly to the scaling relations for displacement thickness and wall shear stress \eqref{eq:delta_corrected}--\eqref{eq:shear_corrected}. 
}

\section{Three-dimensional effects} \label{app:3d}

\begin{figure*}[t!]
    \centering
    (a) \hfill (b) \hfill \hfill \hfill \\
    \includegraphics[trim={0 1.2cm 0 1.4cm},clip,width=0.5\linewidth]{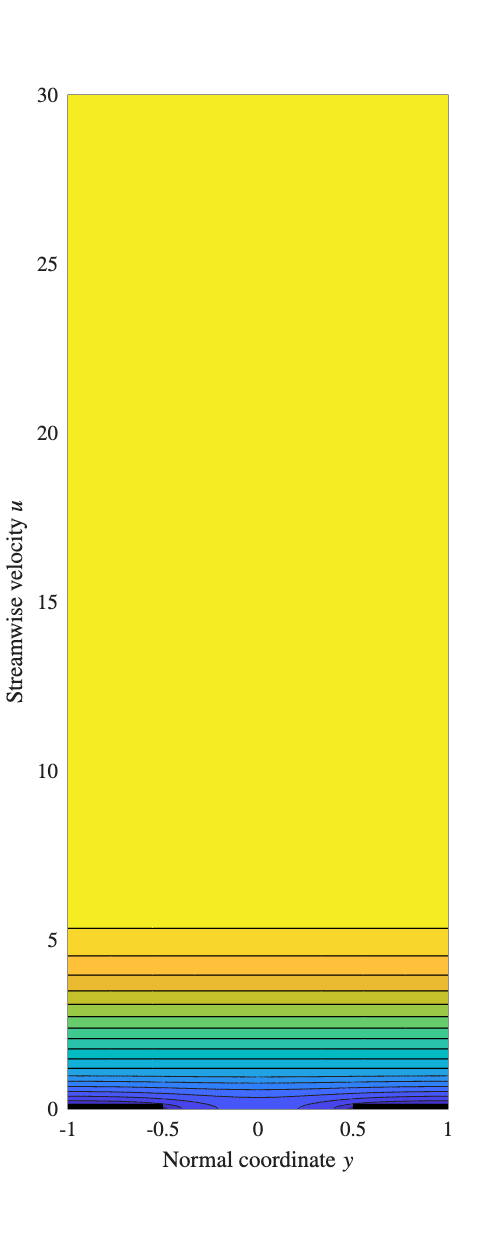}\includegraphics[trim={0 1.2cm 0 1.4cm},clip,width=0.5\linewidth]{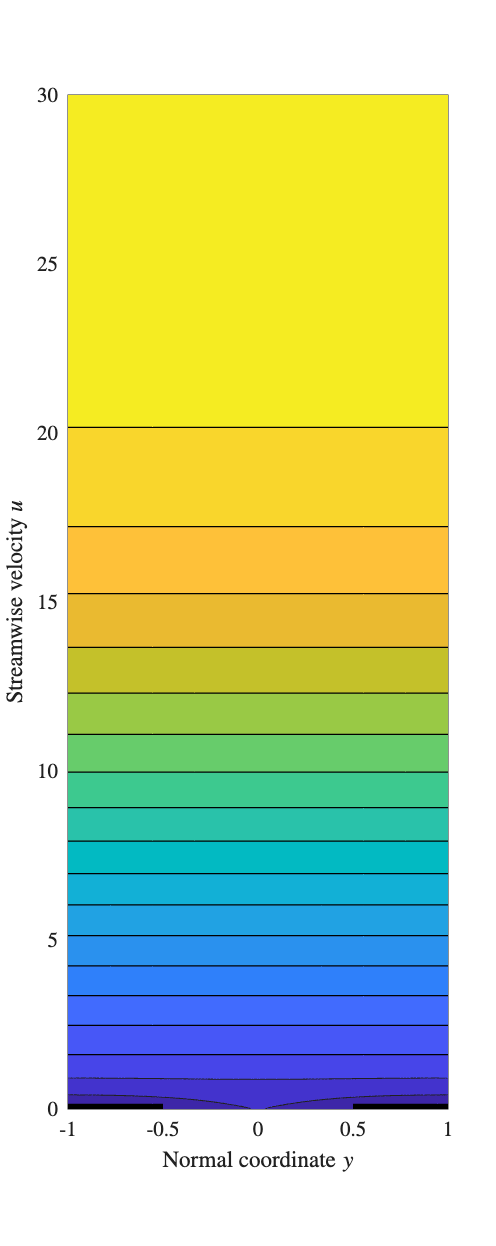} 
    \caption{
    \update{Streamwise velocity $u$ for $\lambda = 1$ at (a) $x = 3$ and (b) $x=30$, for the flow over longitudinally ridged SHSs. 
    The 2D boundary-layer solution (\ref{eq:middle_1}, \ref{eq:middle_1b}, \ref{eq:middle_2}, \ref{eq:matching_1}) is contoured alongside the inner viscous solution for periodic longitudinal ridges from~\cite{philip1972flows}, with streamwise velocities matched at the intermediate scale $y = \epsilon = 1$. 
    }
    }
    \label{fig:7}
\end{figure*}

\update{
The present analysis is restricted to steady, 2D boundary layers. 
Three-dimensionality may enter the problem in two ways through the underlying surface texture or through the boundary-layer flow itself. 
First, the inner (texture-scale) problem may be 3D even when the boundary-layer flow remains 2D. 
For example, longitudinal ridges or riblets lead to inner solutions that depend on both the wall-normal and spanwise coordinates~\cite{philip1972flows}. 
In such cases, homogenisation of the 3D inner problem yields a slip length that depends on the magnitude of the imposed shear, while the outer boundary-layer equations remain 2D. 
The resulting 3D velocity field can then be reconstructed by superposing the periodic inner solution onto the 2D boundary-layer flow through asymptotic matching procedures (see Figure~\ref{fig:7}). 
}

\update{Second, 3D boundary layers arise when the spanwise velocity component is non-zero, for example due to swept geometries or turbo-machinery blades. 
In this case, the boundary-layer equations must be formulated in 3D, and the slip condition becomes tensorial, with distinct streamwise and spanwise slip lengths for SHSs and riblets. 
Under the assumption of a steady, incompressible, zero-pressure-gradient external flow with uniform free-stream velocities, the 3D boundary-layer equations are given by 
\begin{align} 
\label{eq:middle_1_3d}
    u'_{x'} + v'_{y'} + w'_{z'} &= 0, \\
    u' u'_{x'} + v' u'_{y'} + w' u'_{z'} &= u'_{y'y'}, \\ 
    u' v'_{x'} + v' v'_{y'} + w' v'_{z'} &= v'_{y'y'}, \\ 
    u' w'_{x'} + v' w'_{y'} + w' w'_{z'} &= w'_{y'y'},
\end{align}
subject to 
\begin{equation} 
\label{eq:matching_2_3d}
    u' = \lambda_{x'} u'_{y'}, \quad v' = 0, \quad w' = \lambda_{z'} w'_{y'} \quad \text{at} \quad y' = 0, 
\end{equation}
and 
\begin{equation} \label{eq:middle_3_3d}
    u' \to 1, \quad w' \to W', \quad \text{as} \quad y' \to \infty,
\end{equation}
where $W'$ is the ratio of the free-stream spanwise velocity to the free-stream streamwise velocity. 
A complete treatment of such 3D boundary layers, including their linear stability properties, requires a fully 3D numerical formulation and lies beyond the scope of the present work.
}

\backmatter







\subsection*{Declarations}

\bmhead{Funding}  
No specific funding was received for this work.

\bmhead{Competing interests}  
The authors declare that they have no competing interests.

\bmhead{Data availability}  
Data supporting the findings of this study are available from the corresponding author on reasonable request.

\bmhead{Authors' contributions}  
Both authors contributed equally to the conception, analysis and writing of the manuscript.

\bmhead{Acknowledgements}  
SDT acknowledges funding from Research England's E3 fund via the M$^3$4Impact programme. 
DTP was partially supported by the EPSRC under grant EP/V062298/1.

\bibliography{refs}

\end{document}